\documentstyle[aps,prl,preprint,epsfig]{revtex}

\begin{document}
\title{Orbital Magnetoresistance in the La$_{2-x}$Sr$_x$CuO$_4$ System}
\author{F. F. Balakirev$^1$,~~\cite{if}I. E. Trofimov$^1$, S. Guha$^1$,
Marta  Z. Cieplak$^{1,2}$, and P. Lindenfeld$^1$}

\address{$^1$ Department of Physics and Astronomy, Rutgers University,
Piscataway, NJ 08855, USA\\ $^2$ Institute of Physics, Polish Academy
of Sciences, Warsaw, Poland}
\maketitle

\begin{abstract}

Measurements of resistivity, Hall effect, and magnetoresistance have
been made on seven c--axis oriented thin--film specimens of
La$_{2-x}$Sr$_x$CuO$_4$ with values of $x$ from 0.048 to 0.275, and one
specimen that also contains Nd.  The orbital magnetoresistance 
is found not to be proportional to the square of the tangent of the Hall
angle except for values of $x$ near 0.15 above about 80K. 
For smaller values of $x$ the temperature dependence of the magnetoresistance
is different, but quite similar in the various specimens, in spite of
large differences in resistivity, Hall coefficient, and Hall angle.

\end{abstract}

\newpage
\narrowtext


While there is still no consensus on the origin of high--temperature 
superconductivity, it is generally believed that the key to further
understanding is likely to be found in the anomalous properties of the
normal state of these materials. Most notably, the properties seem
not to be in accord with the Landau Fermi--liquid theory of
conventional metals. \cite{pwa,ea} It has been suggested that
the anomalous transport properties may require two distinct scattering
rates for their description. \cite{and}
One of them, $\tau_{tr}^{-1}$, governs conductivity 
and photoemission, and for optimally--doped compounds varies linearly with 
temperature  up to 1000K. \cite{tak}  The other, $\tau_H^{-1}$,
is inferred from the Hall effect and varies as  T$^2$ over a 
broad range of temperatures and carrier concentrations. \cite{chi,xia}
The  orbital magnetoresistance (MR) is an additional property that may
be used to investigate the behavior of the scattering rates, and to
test the hypothesis that there are two separate relaxation times. 

In a normal metal with a single relaxation time and an anisotropic
Fermi surface the weak--field MR, $\Delta\rho/\rho$ = [$\rho(H) -
\rho(0)]/\rho(0)$, 
is proportional to $(\omega_c\tau)^2$, where $\omega_c$ is 
the cyclotron frequency. If the relaxation time is the same as that
which governs the Hall effect, $\Delta\rho/\rho$ is then expected to be
proportional to $(\tan\Theta_H)^2$, where $\Theta_H$ is the Hall angle.
In recent studies this relationship was indeed found to be followed,
by Harris  {\it et al.} \cite{ong}
for two samples of YBa$_2$Cu$_3$O$_{7-\delta}$ (YBCO) and one of
La$_{2-x}$Sr$_x$CuO$_4$ (LSCO) down to about 100K, and by Hussey {\it et
al.} \cite{mck}  in overdoped Tl$_2$Ba$_2$CuO$_6$.

In this letter we report measurements on a series of specimens of the LSCO
system which show that the expected relationship between the magnetoresistance
and the Hall effect is followed over only a rather limited range
of temperature and composition, namely above about 80 K and most closely 
for the optimally doped compound, with departures that grow
as the metal--insulator transition is approached. Although the ratio 
$(\Delta\rho/\rho)/(\tan\Theta_H)^2$ is generally not constant, as would be
expected if the MR and the Hall effect are governed by the same
relaxation time, the functional form of the temperature dependence of
this ratio is similar for all of the underdoped specimens in our study. 

The specimens are c--axis oriented films, about 5000 \AA~thick, made by
pulsed laser deposition on LaSrAlO$_3$ substrates. \cite{cie} The specimen
with $x$ = 0.105 was, in addition, annealed in high--pressure oxygen,
leading to a higher value of T$_c$. \cite{tro} The compositions 
are nominal, as determined from the weights of the materials used for the
targets, but previous work has shown them to be close to the film compositions.
The films were patterned by photolithography, with arms for the longitudinal 
and transverse voltage measurements. Silver pads were evaporated on them
and gold wires attached with silver epoxy or indium solder.

The measurements were made in a cryostat with an 8 T superconducting magnet. 
The temperature was measured with a {\it cernox} thermometer \cite{lak}
and stabilized to about 3 parts in $10^6$ with a computer--controlled
feedback loop.
The magnetoresistance of the sensor was measured separately (see also Ref.
\cite{bra}), and in any case
does not affect the orbital magnetoresistance, which was assumed to be
the difference between the measurements made in the transverse 
(field perpendicular to the current and parallel to the c--axis) and the
longitudinal (field parallel to the current) orientations.

The values of T$_c$ are, in general, somewhat lower than those of bulk single
crystals of the same composition, and the resistivities somewhat higher.
\cite{cie} Both parameters depend sensitively not only on the metal
concentration, but also on disorder and oxygen content, which are more
difficult to control. The metal--insulator transition occurs at about 
$x$ = 0.05, and in its vicinity small changes in composition correspond to
large changes in T$_c$ and $\rho$. In addition the measurement
of the resistivity depends on the thickness, which has an
uncertainty of the order of 20\%.

\begin{figure}[t]
\epsfig{file=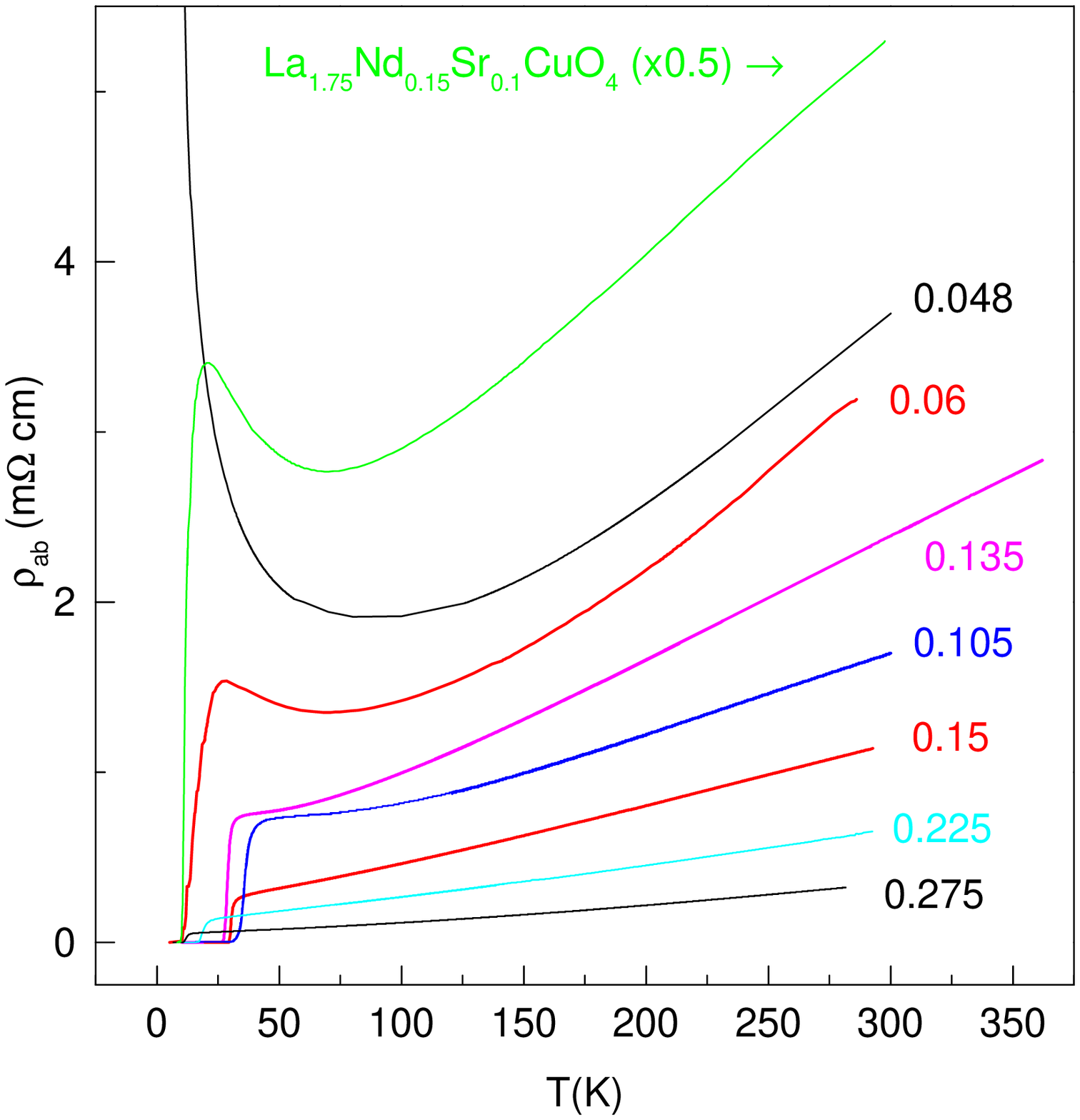,width=0.8\textwidth}
\caption{ 
The resistivity of all samples  as a function of temperature.}
\label{fig1}
\end{figure}

The characteristics of the specimens are shown in Table I. Specimens 1 to 7
differ in their La--Sr ratio, with values of $x$ of 0.048, 0.06, 0.105, 0.135,
0.15, 0.225, and 0.275. Specimen 8 also contains Nd
(La$_{1.75}$Nd$_{.15}$Sr$_{0.1}$CuO$_4$), with a value of T$_c$ 
close to that of specimen 2 ($x$ = 0.06), but with a resistivity larger
by a factor of about three.  Its resistivity has a much more
pronounced maximum near T$_c$, as can be seen on Fig.~\ref{fig1}, which shows
the in--plane resistivity, $\rho$, as a function of temperature 
for all specimens.

Fig.~2 shows the cotangent of the Hall angle, $cot\Theta_H =
(\omega_c\tau_H)^{-1}$, as a function of T$^2$. The measured points
fall on straight lines ($bT^2 + c$) over a substantial temperature
range for the specimens with $x \leq 0.15$, as expected for a relaxation
time proportional to T$^{-2}$. For
the underdoped specimens there is an upturn at low T, in the region where
$\rho$(T) also departs from linearity, as observed previously. \cite{hwa,ken}
The highly overdoped specimen with $x$ = 0.275 exhibits substantial
curvature on this graph, becoming proportional to T$^{1.4}$ in the
high--T region, which is also the temperature dependence of the
resisitivity for this specimen. This behavior
is in keeping with the approach to the normal--metal regime, with
a single scattering time, and adherence to Kohler's rule,
$\Delta\rho/\rho \propto (H/\rho)^2$, as $x$ increases to the strongly
overdoped regime. \cite{kim} Weak curvature is also apparent
for the specimen with $x$ = 0.225. In the opposite regime, for
the specimen with $x$ = 0.048, there is a sharp drop as T goes to zero,
reflecting the divergence of the Hall coefficient at the
metal--insulator transition.

The magnetoresistance is positive and proportional to $H^2$ 
from approximately 40K to 300K. The field dependences of the 
resistances were fitted to parabolas and normalized to 1 tesla. 
The transverse MR increases as $x$ increases, while the longitudinal
MR, which is generally much smaller, decreases. We attribute the
longitudinal MR to isotropic spin scattering, which may be expected to
increase as the antiferromagnetic insulating state is approached,
consistent with our results.

Fig.~3 shows the orbital magnetoresistance, i.e. the difference between the
transverse and the longitudinal MR, as a function of temperature for all
specimens, with the curves for the underdoped specimens and the one
containing neodymium on the lower graph, and the curves for the
optimally doped ($x$ = 0.15) and overdoped specimens on the upper graph.
The figure exhibits several striking and unexpected features. One 
is that the magnitude of the MR does not vary greatly over the whole
range of specimens, especially in the high--T region. Another is
that the curves for the underdoped specimens with $x \leq 0.105$ are
concave over the whole measurement range. This is the opposite
curvature from that expected if the MR is proportional to $\tau_H^2$,
i. e. to (bT$^2$ + c)$^{-2}$.
The curves for the specimens with $x$ = 0.135 and 0.15  show points of
inflection, and those for the overdoped ones are convex, except at the
lowest temperatures.

In Ref. \cite{ong} deviations below 100K
were apparent for all three of their specimens, and were ascribed to
superconducting fluctuations. This was not unreasonable for the two
YBCO specimens with values of T$_c$ of 90K and 60K, and only mildly
surprising for the LSCO specimen with its T$_c$ of 38K. Since,
however, the data do not scale with (T--T$_c$)/T$_c$, and
the specimen with $x$ = 0.048 shows no signs of superconductivity in
the range of the measurements, it is evident that the upturn at the
low--T end of the curves is not caused by superconducting
fluctuations, except, perhaps, to a minor extent.

A particularly interesting case is that of the specimen with strontium content
$x$ = 0.1 and 0.15 neodymium. Its magnetoresistance is almost
indistinguishable from that of the specimen with $x$ = 0.06, which has  
almost the same transition temperature. At the same time the cotangent of
its Hall angle places it close to the optimally doped specimen, except for the
fact that its upturn on Fig.~2 is larger than that of any other
specimen. Its resistivity, the resistance peak at low T, and
Hall coefficient are all larger than those of any other specimen. 
On the other hand R(T) is quite straight above the peak, unlike its
underdoped neighbors on Fig.~1. 
The fact that the behavior of
these quantities does not seem to be correlated with those of the
specimens without Nd emphasizes the earlier conclusion that not only
the resistivity, but also the Hall effect seem to be quite decoupled
from the magnetoresistance. Its carrier concentration and the amount
of disorder
seem to put this specimen in a quite different category from the
others, and a more extended set of measurements with
this and other impurities will be necessary to allow the
effects of these parameters to be assessed and separated.

To emphasize the unexpected relation between Hall effect and MR we show
on Fig.~4 the orbital MR divided by the square of the Hall angle, as a
function of T.  It is apparent that there is no proportionality between the 
magnetoresistance and the square of the Hall angle, except for the
high--temperature part of the curve for the optimally--doped specimen and,
approximately, for $x$ = 0.225. The graph also shows that the
temperature dependences for all of the underdoped specimens are
similar to each other, with a minimum at about 150K.

The point of inflection that was mentioned earlier is most readily apparent
in the MR curves for the specimens with values of $x$ of 0.15 and 0.135. 
It seems also to be present, even if less distinctly, for $x$ = 0.225 and 
$x$ = 0.275, moving to progressively lower T as $x$ increases. The fact
that no point of inflection is seen in the curves for smaller $x$ may 
be because it has then moved to temperatures beyond those of this experiment.
The temperature dependence of the inflection point leads us to suggest
the possibility that it may be correlated with
the opening of a pseudogap in the normal state. \cite{ea,bat,puc}

We now consider some earlier measurements, which are, in general, less precise
than those of Ref.\cite{ong} and those that we present here.
Lacerda et al. \cite{lac}
measured the magnetoresistance of a single crystal with $x$ = 0.075.  They
state that the magnetoresistance is proportional to T$^{-2}$, but their
result is, in fact, in qualitative agreement with ours, considering the larger
uncertainty of their measurement. Preyer et al. \cite{pre} measured single
crystals with $x$ = 0.02, 0.06, and 0.1. They find an isotropic negative
magnetoresistance, and we can only suspect that their specimens contained
unknown magnetic impurities subject to strong spin scattering.

A study by Kimura et al. \cite{kim} describes measurements on a series 
of samples that partially overlap those described here. For their sample
with $x$ = 0.09 they conclude that the magnetoresistance is determined
entirely by superconducting fluctuations up to 100K.
Our work conflicts with this conclusion since the
magnetoresistance does not scale with (T--T$_c$)/T$_c$, as we
stated earlier.
A detailed comparison is made difficult since they do not show 
graphs of $\Delta\rho/\rho$ as a function of T, but there are indications 
of differences from our results. Refs. \cite{ong} and \cite{kim} emphasize
the departures from Kohler's rule, and in this respect our results agree
with theirs.

To summarize, we see that the most interesting and novel aspect of our
results is the description of the MR in the underdoped region
($x \leq 0.135$). Here the temperature dependence of the MR is similar
for all of our specimens, even though there
are large differences in resistivity, Hall coefficient, and Hall angle. 
The temperature dependence of the ratio of the orbital magnetoresistance
to the square of the Hall angle is also similar in these specimens 
(although with different magnitude), with a minimum at about 150K. 
For the optimally doped specimen ($x$ = 0.15) and the next overdoped
specimen ($x$ = 0.225) the proportionality between $\Delta\rho/\rho$
and $\Theta_H^2$ is followed, at least approximately, and for even
higher values of $x$ normal--metal behavior is recovered.
The point of inflection in the temperature dependence of
the MR may be a signal of the opening of a pseudogap.

We would like to thank M. Gershenson for the use of his equipment, and
M. Berkowski for the substrates.  We also thank
E. Abrahams, P. Coleman, G. Kotliar, A. P. Mackenzie, and N. P. Ong 
for helpful discussions. This work was supported, in part, by the
National Science Foundation under grants DMR 93-05860 and DMR
95-01504. M. Z. C. was partly supported by the Polish
Committee for Scientific Research, KBN, under grant 2P03B 05608.

\narrowtext

\begin{figure}[t]
\epsfig{file=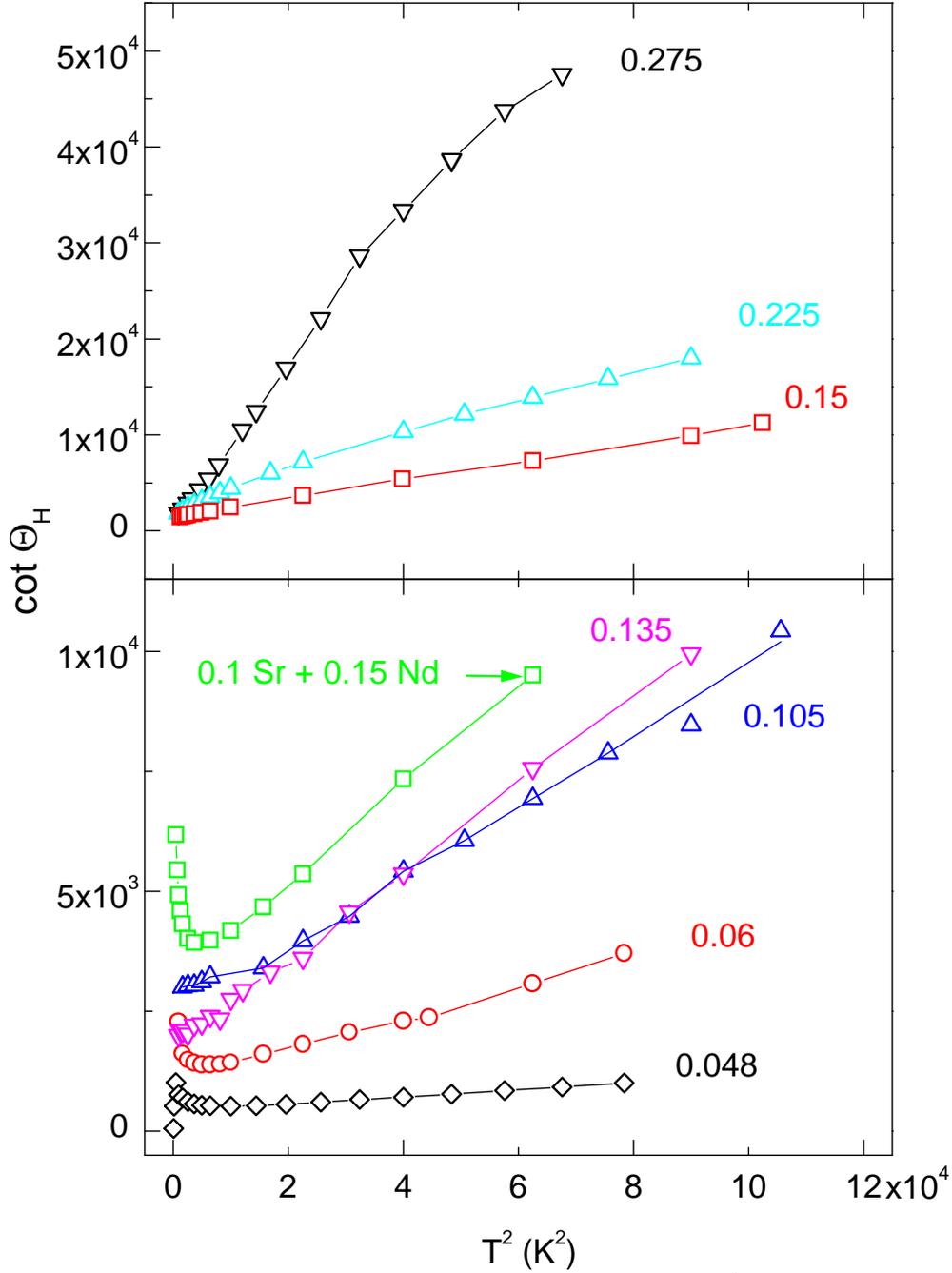,width=0.8\textwidth}
\caption{ 
The cotangent of the Hall angle as a function of T$^2$. 
The straight--line portions were fitted to the relation $bT^2 + c$.
The coefficients $b$ and $c$ are given in Table 1.}
\label{fig2}
\end{figure}

\begin{figure}[t]
\epsfig{file=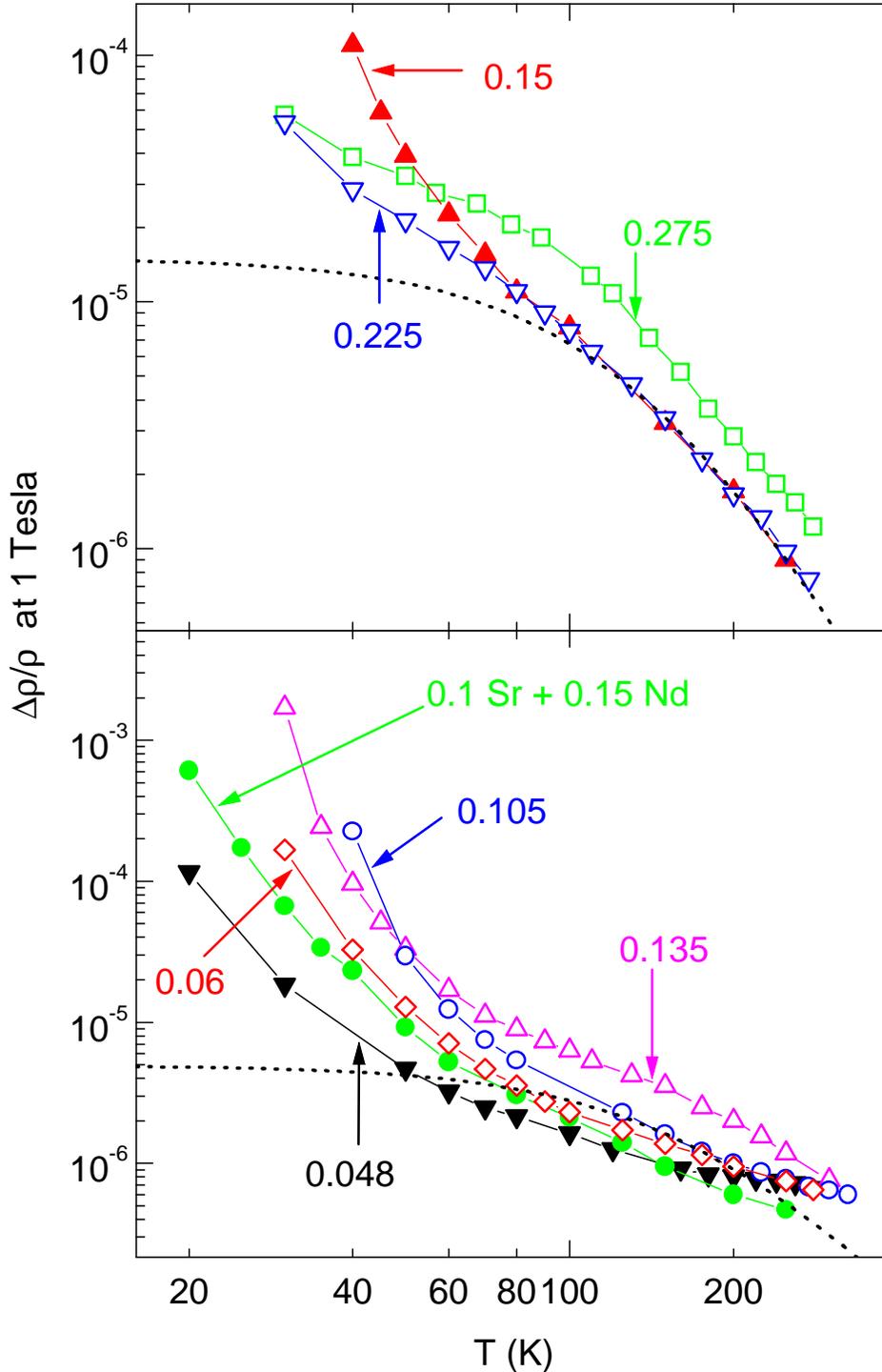,width=0.75\textwidth}
\caption{ 
The orbital magnetoresistance as a function of temperature, for all
specimens, normalized to 1 tesla. 
The dotted lines follow the equation  $a/(b T^2+c)^2$, with the
scaling constant, $a$, chosen to fit the high-temperature data, on the
upper graph for $x$ = 0.15 and on the lower graph for $x$ = 0.105.}
\label{fig3}
\end{figure}

\begin{figure}[t]
\epsfig{file=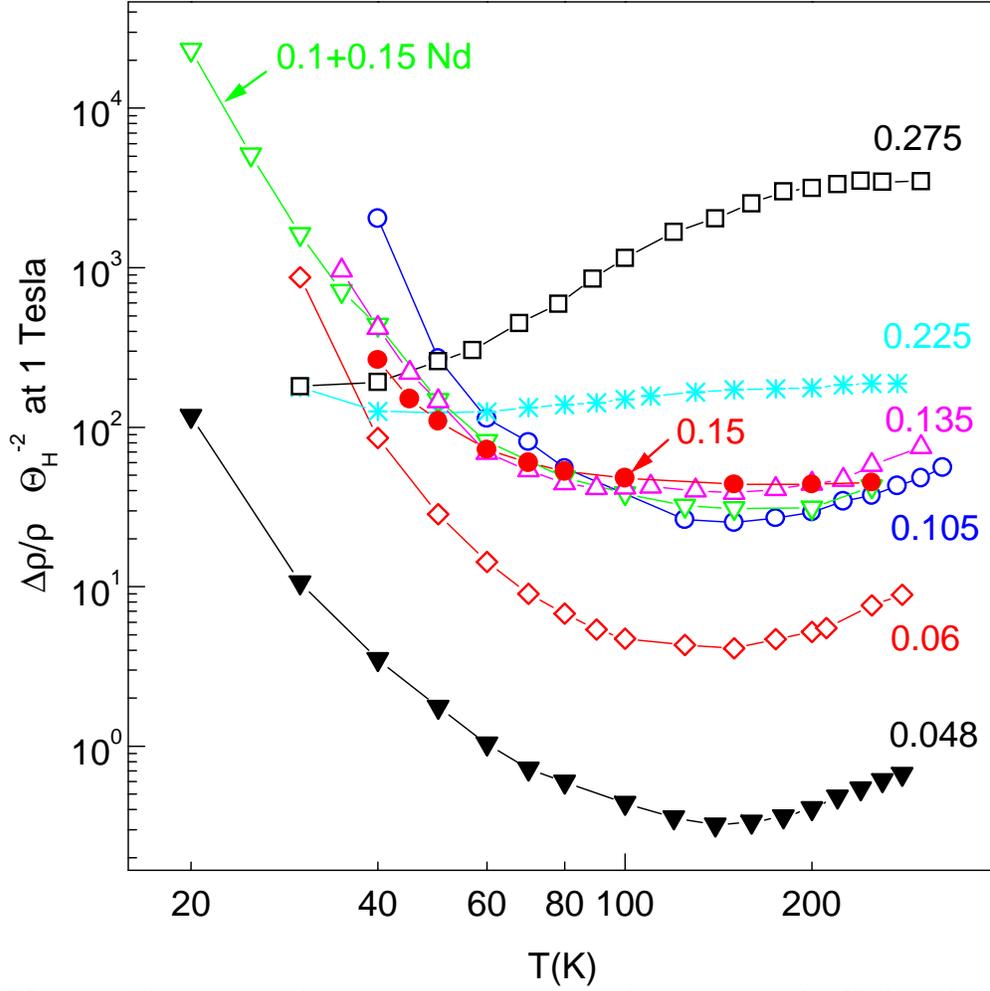,width=0.8\textwidth}
\caption{ 
The ratio of the magnetoresistance to the square of the Hall
angle for all specimens, normalized to 1 tesla.}
\label{fig4}
\end{figure}

\begin{table}

\caption{Characteristics of the specimens,
$La_{2-x-y}Sr_{x}Nd_{y}CuO_{4}$. \\
The coefficients $b$ and $c$ are obtained by fitting the relation
$bT^2 + c$ to the straight--line portions of the Hall effect data on Fig.~2.
}
\begin{tabular}{|c|c|c|c|c|c|c|}
sample   & x     & y    & $T_c$ (K) & $\rho_{300}$ (mOhm cm) &
\mbox{$b$} & $c$ \\
\hline
1        & 0.048 & 0    & N/A   & 3.9 & 0.0075 & 415 \\
\hline
2        & 0.06  & 0    & 10.2  & 3.4  & .031 & 1050 \\
\hline
3        & 0.105 & 0    & 31    & 1.7   & .075 & 2250 \\
\hline
4        & 0.135 & 0    & 27    & 2.3 & .09 & 1825 \\
\hline
5        & 0.15  & 0    & 29.5  & 1 & .095 & 1455 \\
\hline
6        & 0.225 & 0    & 16.5  & 0.67 & N/A & N/A \\
\hline
7        & 0.275 & 0    & 10.7  & 0.345 & N/A & N/A \\
\hline
8        & 0.1   & 0.15 & 9.6    & 10.6  & .103  & 3060 \\
\end{tabular}
\label{table1}
\end{table}


\begin{references}

\bibitem[*]{if}
Present address: Department of Electrical Engineering, Princeton
University, Princeton, NJ 08544.


\bibitem{pwa} 
P. W. Anderson, Science 256, 1525 (1992).

\bibitem{ea}
E. Abrahams, J. Phys. I France 6, 2191 (1996).

\bibitem{and}
P. W. Anderson, Phys. Rev. Letters 67, 2992 (1991).

\bibitem{tak}
H. Takagi, B. Batlogg, H. L. Kao, J. Kwo, R. J. Cava, J. J. Krajewski, and 
W. F. Peck, Jr., Phys. Rev. Letters 69, 2975 (1992).

\bibitem{chi}
T. R. Chien, Z. Z. Wang, and N. P. Ong, Phys. Rev. Letters 67, 2088
(1991).

\bibitem{xia}
G. Xiao, P. Xiong, and M. Cieplak, Phys. Rev. B 46, 8687 (1992).

\bibitem{ong}
J. W. Harris, Y. F. Yan, P. Matl, N. P. Ong, P. W. Anderson, T. Kimura,
and K. Kitazawa, Phys. Rev. Letters 75, 1391 (1995).

\bibitem{mck} N. E. Hussey, J. R. Cooper, J. M. Wheatley, I. R.
Fisher, A.  Carrington, A. P. Mackenzie, C. T. Lin, and O. Milat,
Phys. Rev. Letters 76, 122 (1996).

\bibitem{cie}
M. Z. Cieplak, M. Berkowski, S. Guha, E. Cheng, A. S. Vagelos, D. J. 
Rabinowitz, B. Wu,  I. E. Trofimov, and P. Lindenfeld, Appl. Phys. Lett.
65, 3383 (1994).

\bibitem{tro}
I. E. Trofimov, L. A. Johnson, K. V. Ramanujachary, S, Guha,
M. G. Harrison, M. Greenblatt, M. Z. Cieplak, and P. Lindenfeld,
Appl. Phys. Lett. 65, 2481 (1994).

\bibitem{lak}
Lake Shore Cryotronics, Inc., Westerville, Ohio.

\bibitem{bra} B. L. Brandt and D. W. Liu, preprint.

\bibitem{hwa} H. Y. Hwang, B. Batlogg, H. Takagi, H. K. Kao, J. Kwo, 
R. J. Cava, J. J. Krajewski, and W. F. Peck, Jr., Phys. Rev. Letters 72,
2636, (1994).

\bibitem{ken}
P. Xiong, G. Xiao,  and X. D. Wu, Phys. Rev. 47,5516 (1993),
C. Kendziora, D. Mandrus, L. Mihaly, and L. Forro, Phys. Rev. B 14297
(1992).

\bibitem{kim}
T. Kimura, S. Miyasaka, H. Takagi, K. Tamasaku, H. Eisaki, S. Uchida,
K. Kitazawa, M. Hiroi, M. Sera, and N. Kobayashi, Phys. Rev. B 53, 8733 (1996).

\bibitem{bat}
B. Batlogg, H. Y. Hwang, H. Takagi, R. J. Cava, H. L. Kao, and J. Kwo,
Physica C 235-240, 130 (1994).

\bibitem{puc}
A. V. Puchkov, D. N. Basov, and T. Timusk, J. Phys.: Condensed Matter
8, 10049 (1996).

\bibitem{lac}
A. Lacerda, J. P. Rodriguez, M. F. Hundley, Z. Fisk, P. C. Canfield, 
J.D. Thompson, and S. W. Cheong, Phys. Rev. B 49, 9097 (1994).

\bibitem{pre}
N. W. Preyer, M. A. Kastner, C. Y. Chen, R. J. Birgenau, and Y. Hidaka,
Phys. Rev. B 44, 407 (1991).


\end{references}
\end{document}